\title{Broadcasting with side information}
\author{ {Noga Alon\thanks{Schools of
Mathematics and Computer Science, Raymond and Beverly Sackler
Faculty of Exact Sciences, Tel Aviv University, Tel Aviv, 69978,
Israel and IAS, Princeton, NJ 08540, USA. Email: nogaa@tau.ac.il.
Research supported in part by a USA-Israeli BSF grant, by the Israel
Science Foundation and by the Hermann Minkowski Minerva Center for
Geometry at Tel Aviv University.}} \quad {Avinatan
Hasidim\thanks{School of Computer Science, Hebrew University Email:
avinatan@cs.huji.ac.il.}} \quad {Eyal Lubetzky\thanks{Microsoft
Research, One Microsoft Way, Redmond, WA 98052-6399, USA. Email:
eyal@microsoft.com. Research partially supported by a Charles Clore
Foundation Fellowship.}} \quad {Uri Stav\thanks{ School of Computer
Science, Raymond and Beverly Sackler Faculty of Exact Sciences, Tel
Aviv University, Tel Aviv, 69978, Israel. Email:
uristav@tau.ac.il.}}\quad {Amit Weinstein\thanks{ School of Computer
Science, Raymond and Beverly Sackler Faculty of Exact Sciences, Tel
Aviv University, Tel Aviv, 69978, Israel. Email: amitw@tau.ac.il.}}}

\documentclass [11pt]{article}
\usepackage[]{amsmath,amssymb,amsfonts,latexsym,amsthm,enumerate,url}
\usepackage[colorlinks=true, pdfstartview=FitV, linkcolor=blue,
  citecolor=blue, urlcolor=blue]{hyperref}

\date{}

\newtheorem{theorem}{Theorem}[section]
\newtheorem{lemma}[theorem]{Lemma}
\newtheorem{claim}[theorem]{Claim}
\newtheorem{definition}{Definition}

\newtheorem{corollary}[theorem]{Corollary}

\newtheorem{observation}[theorem]{Observation}

\renewcommand{\epsilon}{\varepsilon}

\newcommand{\mais}{\ensuremath\textrm{MAIS}}

\newcommand{\rank}{\operatorname{rank}}

\newcommand{\orprod}{\ensuremath{\mathaccent\cdot\vee}}

\newcommand{\conf}{\mathfrak{C}}
\newcommand{\chif}{\chi_f}

\newcommand{\Z}{\mathbb{Z}}

\newtheoremstyle{upright}%
        {8pt plus2pt minus4pt}%
        {8pt plus2pt minus4pt}%
        {\upshape}%
        {}%
        {\bfseries}%
        {:}%
        {1em}%
        {}%
\theoremstyle{upright}
\newtheorem{remark}[theorem]{Remark}

\newcommand{\ignore}[1]{}

\begin{document}
\maketitle

\begin{abstract}

A sender holds a word $x$ consisting of $n$ blocks $x_i$, each of
$t$ bits, and wishes to broadcast a codeword to $m$ receivers,
$R_1,...,R_m$. Each receiver $R_i$ is interested in one block, and
has prior side information consisting of some subset of the other
blocks. Let $\beta_t$ be the minimum number of bits that has to be
transmitted when each block is of length $t$, and let $\beta$ be the
limit $\beta = \lim_{t \rightarrow \infty} \beta_t/t$. In words,
$\beta$ is the average communication cost per bit in
each block (for long blocks).
Finding the coding rate $\beta$, for such an informed
broadcast setting, generalizes several coding theoretic parameters
related to Informed Source Coding on Demand, Index Coding and
Network Coding.

In this work we show that usage of large data blocks may strictly
improve upon the trivial encoding which treats each bit in the block
independently. To this end, we provide general bounds on $\beta_t$,
and prove that for any constant $C$ there is an explicit broadcast
setting in which $\beta = 2$ but $\beta_1 > C$. One of these
examples answers a question of \cite{LS}.

In addition, we provide examples with the following counterintuitive
direct-sum phenomena. Consider a union of several mutually
independent broadcast settings. The optimal code for the combined
setting may yield a significant saving in communication over
concatenating optimal encodings for the individual settings. This
result also provides new non-linear coding schemes which improve
upon the largest known gap between linear and non-linear Network Coding,
thus improving the results of \cite{DFZ}.

The proofs are based on a relation between this problem and
results in the study of Witsenhausen's rate, OR graph products,
colorings of Cayley graphs, and the chromatic numbers of Kneser
graphs.

\end{abstract}

\newpage

\section{Introduction}

Source coding deals with a scenario in which a \emph{sender} has
some data string $x$ he wishes to transmit through a broadcast
channel to \emph{receivers}. In this paper we consider a variant of
source coding which was first proposed by Birk and Kol \cite{BK}. In
this variant, called Informed Source Coding On Demand (ISCOD), each
receiver has some prior side information, comprising a part of the
input word $x$. The sender is aware of the portion of $x$ known to
each receiver. Moreover, each receiver is interested in just part of
the data.

We formalize this source coding setting as follows. Suppose that a
sender $S$ wishes to broadcast a word $x=x_1x_2\ldots x_n$, where
$x_i \in \{0,1\}^t$ for all $i$, to $m$ receivers $R_1,\ldots,R_m$.
Each $R_j$ has some prior side information, consisting of some of
the blocks $x_i$, and is interested in a single block $x_{f(j)}$.
The sender wishes to transmit a codeword that will enable each and
every receiver $R_j$ to reconstruct its missing block $x_{f(j)}$
from its prior information. Let $\beta_t$ denote the minimum
possible length of such a binary code. Our objective in this paper
is to study the possible behaviors of $\beta_t$, focusing on the
more natural scenario of transmitting large data blocks (namely a
large $t$).

The motivation for informed source coding is in applications such as
Video on Demand. In such applications, a network, or a satellite,
has to broadcast information to a set of clients. During the first
transmission, each receiver misses a part of the data. Hence, each
client is now interested in a different (small) part of the data,
and has a prior side information, consisting of the part of the data
he received \cite{YZ}. Note that our assumption that each receiver
is interested only in a single block is not necessary. To see this,
one can simulate a receiver interested in $r$ blocks by $r$
receivers, each interested in one of these blocks, and all having
the same side information.

The problem above generalizes the problem of Index Coding, which was
first presented by Birk and Kol \cite{BK}, and later studied by
Bar-Yossef, Birk, Jayram and Kol \cite{BBJK} and by Lubeztky and Stav \cite{LS}.
 Index Coding is equivalent to a special case of our
problem, in which $m=n$, $f(j)=j$ for all $j \in [m]=\{1,\ldots,m\}$
and the size of the data blocks is $t=1$. Our work can also be
considered in the context of Network Coding, a term which was coined
by Ahlswede, Cai, Li, and Yeung \cite{ACLY}. In a Network Coding
problem it is asked whether a given communication network (with
limited capacities on each link) can meet its requirement, passing a
certain amount of information from a set of source vertices to a set
of targets.

It will be easier to describe our source coding problems in terms of
a certain hypergraph.  We define a {\em directed hypergraph}
$H=(V,E)$ on the set of vertices $V=[n]$. Each vertex $i$ of $H$
corresponds to an input block $x_i$. The set $E$ of $m$ edges
corresponds to the receivers $R_1,\ldots,R_m$. For the receiver
$R_j$, $E$ contains a directed edge $e_j = (f(j) , N(j))$, where
$N(j) \subset [n]$ denotes the set of blocks which are known to
receiver $R_j$. Clearly the structure of $H$ captures the definition
of the broadcast setting. We thus denote by $\beta_t(H)$ the minimal
number of bits required to broadcast the information to all the
receivers when the block length is $t$.

Let $H$ be such a directed hypergraph. For any pair of integers
$t_1$ and $t_2$, when the block length is $t_1+t_2$, it is possible
to encode the first $t_1$ bits, then separately encode the remaining
$t_2$ bits. By concatenating these two codes we get
$\beta_{t_1+t_2}(H) \leq \beta_{t_1}(H) + \beta_{t_2}(H)$, i.e.
$\beta_t(H)$ is sub-additive. Fekete's Lemma thus implies that the
limit $\lim_{t \rightarrow \infty }\beta_t(H)/t$ exists and equals
$\inf_t \beta_t(H) / t$. We define $\beta(H)$ to be this limit:
$$\beta(H):=\lim_{t \rightarrow \infty }\frac{\beta_t(H)}{t}=
\inf_t \frac{\beta_t(H)}{t}~.$$ In words, $\beta$ is the average
asymptotic number of encoding bits needed per bit in each input block.

To study this problem, we will also consider the following
related one. Let $k \cdot H$ denote the disjoint union of $k$ copies of
$H$. Define $\beta^*_t(H) := \beta_1 (t \cdot H)$. In words,
$\beta_t^*$ represents the minimal number of bits required if the
network topology is replicated $t$ independent times\footnote{Such
a scenario can occur when the topology is standard (e.g. resulting
from using a common application or operation system). Therefore it
is identical across networks, albeit with different data.}. A
similar sub-additivity argument justifies the definition of the limit $$\beta^*(H)
:= \lim_{t \rightarrow \infty} \frac{\beta^*_t(H)}{t} = \inf_t
\frac{\beta^*_t(H)}t~.$$ By viewing each receiver in the broadcast
network as $t$ receivers, each interested in a
single bit, we can compare this scenario with the setting of independent
copies. Clearly, the receivers in the first scenario
have additional information
and hence $\beta_t(H) \leq
\beta^*_t(H)$ for any $t$. Taking limits we get $\beta(H) \le \beta^*(H)$.

There are several lower bounds for $\beta(H)$. One such simple
bound, which we denote by $\vec{\alpha}(H)$, is the maximal size of
a set $S$ of vertices satisfying the following: For every $v\in S$
there exists some $e = (v,J)\in E$ so that $J \cap S = \emptyset$. A
simple counting argument shows that $\vec{\alpha}(H) \le \beta(H) $,
giving\footnote{The bound $\vec{\alpha}$ given here generalizes the
bound given in \cite{BBJK} to directed hypergraphs, as well as to
$\beta$ (rather than just to $\beta_1$). Another bound in the Index
Coding model is the $\mais$ (maximum acyclic induced subgraph) bound
given in \cite{BBJK}, that can also be generalized to our model.}
\begin{equation}\label{eq:beta}\vec{\alpha}(H) \le \beta(H) \le
\beta^*(H) \le \beta_1(H)~.\end{equation}

\subsection{Our Results}

Let $H = ([n], E)$ be a directed hypergraph for a broadcast network,
and set $t=1$. It will be convenient to address the more precise
notion of the \emph{number of codewords} in a broadcast code which
satisfies $H$. We say that $\mathcal{C}$, a broadcast code for $H$,
is \emph{optimal}, if it contains the minimum possible number of
codewords (in which case, $\beta_1(H) = \lceil \log_2
|\mathcal{C}|\rceil$). We say that two input-strings $x,y \in
\{0,1\}^n$ are confusable if there exists a receiver $ e=(i,J) \in
E$ such that $x_i \neq y_i$ but $x_j = y_j$ for all $j \in J$. This
implies that the input-strings $x,y$ can not be encoded with the
same codeword. Denoting by $\gamma$ the maximal cardinality of a set
of input-strings which is pairwise unconfusable. The first technical
result of this paper relates $\beta^*$ and $\gamma$.

\begin{theorem}\label{witsen-thm-1}
Let $H$ and $\gamma$ be
defined as above. The following holds for any integer $k$:
\begin{equation}
\label{witsen-eqcode-bound} \left(\frac{2^n}{\gamma}\right)^k \leq
|\mathcal{C}|
   \leq \Big\lceil \left(\frac{2^n}{\gamma}\right)^k k n
\log 2 \Big\rceil  ~,\end{equation}
where $\mathcal{C}$ is an optimal code for $k \cdot H$. In particular,
$\beta^*(H)=\displaystyle{\lim_{k\to\infty} \frac{\beta_1(k \cdot H)}{k} = n -
\log_2 \gamma}$.
\end{theorem}
A surprising corollary of the above theorem is that $\beta^*$ may be
strictly smaller than $\beta_1$. Indeed, as $\beta^*$ deals with the
case of disjoint instances, it is not intuitively clear that this
should be the case: one would think that there can be no room for
improving upon $\beta_1(H)$ when replicating $H$ into $t$ disjoint
copies, given the total independence between these copies (no
knowledge on blocks from other copies, independently chosen inputs).
Note that even in the somewhat related Information Theoretic notion
of the Shannon capacity of graphs (corresponding to
channel coding rather than source coding), though it is known that the
capacity of a disjoint union may exceed the sum of the individual capacities
(see \cite{NogaUnion}), it is easy to verify that disjoint unions of
the \emph{same graph} can never achieve this. The following theorem
demonstrates the possible gap between $\beta_1(t\cdot H)/t$ and
$\beta_1(H)$ even in a very limited setting, which coincides with
Index Coding. This solves the open problem presented by \cite{LS}
already for the smallest possible $n=5$.
\begin{theorem}\label{witsen-thm-k-C5}
Define a broadcast network $H=\left(\Z_5 ,E\right)$ based on the odd cycle $C_5$:
For each $i \in \Z_5$, there is a directed edge
$\left(i, \left\{ i-1, i+1\right\}\right)$, where the arithmetic is modulo 5. Then $\beta_1(H) = 3$, whereas $\beta^*(H) =  5-\log_2 5
\approx 2.68$.
\end{theorem}
It is worth noting that in the example above, the optimal code for
$H$ contains $8$ codewords, whereas in the limit, each disjoint copy
of $H$ costs $6.4$ codewords, hence this surprising direct-sum
phenomenon carries beyond any integer rounding issues. In addition,
in Section \ref{sec:disjoint_union} (Theorem
\ref{thm-odd-cycle-complements}) we generalize the above example to
any broadcast network which is based on a complement of an odd
cycle.

The following theorem extends the above results on the gap between $\beta^*$ and $\beta_1$ even further, by providing an example where $\beta^*$ is bounded whereas $\beta_1$ can be arbitrarily large:

\begin{theorem} \label{thm-dif-beta-star-beta-1}
There exists an explicit infinite family of broadcast networks for which
$\beta^*(H)<3$ is bounded and yet $\beta_1(H)$ is unbounded.
\end{theorem}

Finally, recalling \eqref{eq:beta}, one
would expect that in many cases $\beta$ should be strictly smaller than $\beta^*$,
as the receivers possess more side
information. However it is not clear how much can be gained by this
additional information. We construct an example where not only is there
a difference between the two, but $\beta$ is constant while
$\beta^*$ is unbounded.


\begin{theorem} \label{thm-dif-beta-beta-star}
There exists an explicit infinite family of broadcast networks for which
$\beta(H)=2$ is constant whereas $\beta^*(H)$ is unbounded.
\end{theorem}



%

We discuss applications of the results to Network and Index coding
in what follows.


\subsection{Related Work}

Our work is a generalization of Index Coding, which was first
studied by Birk  and Kol \cite{BK}. This problem deals with a
sender, who wishes to send $n$ blocks of data to $n$ receivers,
where each receiver knows a subset of the blocks, and is interested
in a single block (different receivers are interested in different
blocks). The sender can only utilize a broadcast channel, and we wish
 to minimize the number of bits he has to send. Birk and Kol
presented a class of encodings, based on erasure Reed Solomon codes.
They also dealt with some of the practical issues of this scheme,
such as synchronization between the clients and the server. Finally
they gave examples for scenarios where their codes were not optimal,
and presented the question of finding better codes. The first
improvement to the original codes was done by Birk, Bar-Yossef,
Jayram and Kol, who found a lower bound to the minimal length of
linear codes, called the \emph{min-rank}. They also conjectured that linear
codes are optimal for index coding, a conjecture that was later
refuted by Lubetzky and Stav. However, the proof by Lubeztky and
Stav was limited, in the sense that they constructed an
index coding problem, for which linear codes over any field
were not optimal,
and yet a combination of linear codes over several
fields may well be optimal.
Theorem \ref{witsen-thm-k-C5} refutes the conjecture in a stronger
sense, by showing an index coding problem for which the optimal
solution is not linear for any field size or even any
combination of several fields.

Network Coding deals with a scenario in which several sources wish
to pass information to several targets, when the communication
network is modeled by a graph. Each edge has a capacity, and the
goal is to see if the network is satisfiable, i.e. if it is possible
to meet all the demands of the clients. This very intuitive model of
communication is motivated by the Internet, where routers pass
information from different sites to users. It has been believed that
routers need only store and forward data (Multi Commodity Flow),
without processing it at all. This intuition was proved false by
Ahlswede, Cai, Li, and Yeung \cite{ACLY}, who showed a very simple
network (the Butterfly Network), which was only satisfiable if one
of the nodes processed the data which entered it. The encoding done
in this example was linear, and for some time it was not clear if
non linearity is beneficial in constructing optimal network codes.
The work of Dougherty, Freiling, and Zeger \cite{DFZ} answered this
in the affirmative, giving an example of a network in which non
linear codes are essential in order to achieve the required network
capacity. Their construction relies on the parity of the
characteristic of the underlying field, and gives a ratio of $1.1$
between the coding capacity and the linear coding capacity. Another
way to achieve a gap between linear and non linear codes was
presented in \cite{CG}. Improving this ratio, as well as finding new
ways to create such gaps are open problems in the field of Network
Coding (see \cite{YLCZ} for a survey).

To see that our model is indeed a special case of network coding, we
present the following simple
reduction between a directed hypergraph which describes
a broadcast network $H=(V,E)$ to a network coding problem. We build
a network of $n$ sources $s_1, \ldots, s_n$, and $m$ sinks $t_1,
\ldots, t_m$. There are also two special vertices $u$ and $w$.
Letting
\begin{align*}
E_{\infty} = \{(s_i, u) : i \in [n]\} ~ \cup~ \{(w, t_e):  e \in E\} ~ \cup~
  \{(s_j, t_e): e= (i,J) \in E, j \in J \}~,
 \end{align*} the network has an edge with infinite capacity for each $e \in
E_{\infty}$. In addition to that, the network has a single edge with
finite capacity, $(u,w)$. If each source receives an input of $t$
bits, the demand of the network can be satisfied if and only if the
capacity of $(u,w)$ is at least $\beta_t(H)$. Moreover, this
reduction maintains linearity of codes.

This reduction enables us to translate some of our results to the
network coding model. In particular, we prove the following
corollary of our results,  improving the results of
\cite{DFZ}:

\begin{corollary}\label{network-code-ratio}
There exists a network with 48 vertices such that the ratio between
the coding capacity and the linear coding capacity in it is at least
$1.324$.
\end{corollary}

The corollary is based on the results of Appendix
\ref{sec:odd_holes_numbers}, where we show that for a certain directed
hypergraph, $H_{23}$, any linear code requires $3$ bits,
while $\beta^*(H_{23}) \leq 2.265$.

\section{Optimal codes for a disjoint union of directed hypergraphs}\label{sec:witsen-sec-main}

The size of an optimal code for a given directed hypergraph
describing a broadcast network may be restated as a problem of
determining the chromatic number of a graph, as observed by Bar-Yossef
et al. for the Index Coding model \cite{BBJK}. Consider the block-length $t=1$, and define the following:


\begin{definition}[\textbf{Confusion graph}] Let $H=([n],E)$ be a directed
hypergraph describing a broadcast network. The \emph{confusion graph} of $H$,
$\conf(H)$, is the undirected graph on the vertex set $\{0,1\}^n$,
where two vertices $x\neq y$ are adjacent iff for some
$e = (i,J) \in E$, $x_i \neq y_i$ and yet $x_j=y_j$ for all $j\in J$.
\end{definition}

In other words, $\conf(H)$ is the graph whose vertex set is all
possible input-words, and two vertices are adjacent iff they are confusable,
meaning they cannot
be encoded by the same codeword for $H$ (otherwise,
the decoding of at least one of the receivers would be ambiguous).
Hence, a code for $H$ is equivalent to a legal vertex
coloring of $\conf(H)$, where each color class corresponds to a
distinct codeword. Consequently, if $\mathcal{C}$ is an optimal
code for $H$, then $|\mathcal{C}| = \chi(\conf(H))$.

Similarly, one can define $\conf_t(H)$, the confusion graph corresponding to $H$ with block-length $t$.
From now on, throughout this section, the length $t$ of the blocks considered will be $1$.

\begin{proof}[\textbf{\emph{Proof of Theorem \ref{witsen-thm-1}}}]
The OR graph product is equivalent to the complement of the
\emph{strong product}\footnote{Namely, the OR product of $G_1$ and $G_2$
is the complement of the strong product of $\overline{G_1}$ and
$\overline{G_2}$.}, which was thoroughly studied in the investigation
of the Shannon capacity of a graph, a notoriously challenging graph
parameter introduced by Shannon \cite{Shannon}.
\begin{definition}[\textbf{OR graph product}]
The \emph{OR graph product} of $G_1$ and $G_2$, denoted by $G_1 \orprod
G_2$, is the graph on the vertex set $V(G_1) \times V(G_2)$, where $(u,v)$
and $(u',v')$ are adjacent iff either $u u' \in E(G_1)$ or $v v' \in
E(G_2)$ (or both). Let $G^{\orprod k}$ denote the $k$-fold OR product
of a graph $G$.
\end{definition}
%

Let $H_1$ and $H_2$ denote directed hypergraphs (as before) on the
vertex-sets $[m]$ and $[n]$ respectively, and consider an encoding
scheme for their disjoint union, $H_1+H_2$. As there are no edges
between $H_1$ and $H_2$, such a coding scheme cannot encode two
input-words $x,y\in\{0,1\}^{m+n}$ by the same codeword iff this
forms an ambiguity either with respect to $H_1$ or with respect to
$H_2$ (or both). Hence:
\begin{observation}
For any pair $H_1,H_2$ of directed hypergraphs as above, the graphs
$\conf(H_1+H_2)$ and $\conf(H_1) \orprod \conf(H_2)$ are isomorphic.
\end{observation}
Thus, the number of codewords in an optimal  code for $k \cdot
H$ is equal to $\chi(\conf(H)^{\orprod k})$. The chromatic numbers
of strong powers of a graph, as well as those of OR graph powers,
have been studied intensively. In the former case, they correspond to
the Witsenhausen rate of a graph (see \cite{Witsenhausen}). In the
latter case, the following was proved by McEliece and Posner
\cite{MP}, and also by Berge and Simonovits
\cite{BS}:
\begin{equation} \label{witsen-eq-berge-simonovits}
\lim_{k\to\infty}\left(\chi(G^{\orprod k})\right)^{1/k} =
\chif(G)~,
\end{equation}
where $\chif(G)$ is the \emph{fractional chromatic number} of the
graph $G$, defined as follows. A legal vertex coloring corresponds
to an assignment of $\{0,1\}$-weights to independent-sets, such that
every vertex will be ``covered" by a total weight of at least $1$. A
fractional coloring is the relaxation of this problem where the
weights belong to $[0,1]$, and $\chif$ is the minimum possible sum
of weights in such a fractional coloring.

To obtain an estimate on the rate of the convergence in
\eqref{witsen-eq-berge-simonovits}, we will use the following
well-known properties of the fractional chromatic number and OR
graph products (cf.
\cite{NogaRepeatedComm},\cite{LovaszChi},\cite{LinialVazirani} and
also \cite{Feige}):
\begin{enumerate}[(i)]
  \item For any graph $G$, $\chif(G^{\orprod k}) =
      \chif(G)^k$.
  \item \label{witsen-eq-chi-bound} For any graph $G$,
      $\chif(G) \leq \chi(G) \leq \lceil \; \chif(G) \log
      |V(G)| \; \rceil$. [This is proved by selecting $\lceil \; \chif(G) \log
      |V(G)| \; \rceil$ independent
      sets, each chosen randomly and independently according to
      the weight distribution, dictated by the optimal
      weight-function achieving $\chif$. One can show that the expected number of uncovered vertices is less than
      $1$.]
  \item For any vertex transitive graph $G$ (that is, a
      graph whose automorphism group is transitive), $\chif(G) = |V(G)| /
      \alpha(G)$ (cf., e.g., \cite{AlgabraicGraphTheory}), where $\alpha(G)$ is the independence number of $G$.
\end{enumerate}
In order to translate \eqref{witsen-eq-chi-bound} to the statement
of \eqref{witsen-eqcode-bound}, notice that $\gamma$, as defined in
Theorem \ref{witsen-thm-1} is precisely $\alpha(\conf(H))$,
the independence number of the confusion graph. In addition, the
graph $\conf(H)$ is indeed vertex transitive, as it is a Cayley
graph of $\Z_2^n$. Combining the above facts, we obtain that:
$$\chif\left(\conf(H)^{\orprod k}\right)^{1/k} = \frac{2^n}{\alpha(\conf(H))} =
\frac{2^n}{\gamma}~.$$ Plugging the above equation into
\eqref{witsen-eq-chi-bound}, while recalling that
$\chi\left(\conf(H)^{\orprod k}\right)$ is the size of the optimal
code for $k \cdot H$, completes the proof of the theorem.
\end{proof}
\begin{remark}\label{witsen-rem-lov}
  The right-hand-side of \eqref{witsen-eqcode-bound} can be replaced
  by $\left( \frac{2^n}{\gamma} \right) ^k \lceil 1 + k \log \gamma \rceil$. To see this, combine the
  simple fact that $\alpha(G^{\orprod k}) = \alpha(G)^k$ with the
  bound $\chi(G) \leq \lceil \chif(G) (1 + \log
  \alpha(G))\rceil $ given in \cite{LovaszChi}
  (which can be proved by choosing $\lceil \chif(G) \log \alpha(G) \rceil$ independent sets randomly as before, leaving at most $\lceil \chif(G) \rceil$ uncovered vertices,
  to be covered separately).
\end{remark}

\section[The possible gaps between the three parameters]{The possible gaps between the parameters $\beta$, $\beta^*$ and $\beta_1$} \label{sec:gaps}

As noted in \eqref{eq:beta}, $\beta \leq \beta^* \leq \beta_1$. In
this section, we describe networks where the gap between any two of
these parameters can be unbounded. The first construction is of a
directed hypergraph on $n$ vertices where $\beta = 2$ while
$\beta^*$ and $\beta_1$ are both $\Theta(\log n)$, for any $n=2^k$.
We then describe a more surprising, general construction which
provides a family of directed hypergraphs, for which $\beta^* < 3$
while $\beta_1 = \Theta( \log\log n)$. Finally, we describe
simple scenarios where even in the restricted Index Coding model,
taking disjoint copies of the network can be encoded strictly better
than concatenating the encodings of each of the copies. These
constructions also apply to network coding, where for the latter
ones it is also possible to prove a lower bound on the length of the
optimal linear encoding scheme.

Throughout this section, we use the following notations. For binary vectors $u$ and $v$, let
$|u|$ denote the Hamming weight of $u$, and let $u\oplus v$ be the
bitwise xor of $u$ and $v$.

\subsection[Proof of Theorem 1.4]{Proof of Theorem \ref{thm-dif-beta-beta-star}} \label{sec:gap_beta_beta_star}

%
Consider a scenario in which the input word consists of a block
$x_i$ for each nonzero element $i \in \Z_2^k$, thus the number of blocks
is $n=2^k-1$. For any pair of distinct nonzero elements $i,j
\in \Z_2^k$, there exists a receiver which is interested in the block
$x_i$ and knows all other blocks except for $x_j$. This scenario
corresponds to a directed hypergraph $H$ in which the vertices are
the nonzero elements of $\Z_2^k$, and for every $i,j \in \Z_2^k$ we have a
directed edge $(i, \Z_2^k \setminus \{0,i,j\})$.

Let $\conf=\conf(H)$ be the confusion graph of $H$ for
block-length $t=1$. Since each receiver is missing precisely two
blocks, for any pair of distinct codewords $u,v \in \{0,1\}^n$, whenever $|u
\oplus v| \geq 3$, every receiver can distinguish between the
codewords. On the other hand, if $|u \oplus v| \leq 2$, there is
some receiver who may confuse his block in $u$ and $v$. Thus,
$\conf$ is exactly the Cayley graph of $\Z_2^n$ whose generators are
all elementary unit vectors $e_i$, as well as all vectors $e_i \oplus e_j$.

\begin{claim} The above defined graph $\conf$ satisfies $\chi(\conf) \leq n+1 $.
\end{claim}

\begin{proof}Since $n=2^k-1$ there is a Hamming code of
length $n$, and the required coloring of any vector $u$ is  given by
computing its syndrome.
More explicitly,
define $c:V(\conf)
\mapsto \Z_2^k$ as follows. For a vector $u=(u_1,
\ldots ,u_n) \in \Z_2^n$, put $c(u) := \sum_{i \in
\Z_2^k-\{0\}} u_i \cdot i~ $.  Let $u,v \in \Z_2^n$ be a pair of adjacent
vertices in $\conf$. Thus  $1 \leq |u \oplus v| \leq 2 $,
and $c(u) \oplus c(v)$ is a sum (in $\Z_2^k$) of one or two distinct nonzero
elements of $\Z_2^k$, which is not zero. Thus $c(u) \neq c(v)$, as needed.
\end{proof}

\begin{claim}The above defined graph $\conf$ satisfies  $\chif(\conf) \geq n + 1 $, and thus $\chi(\conf)=\chif(\conf)
=n+1$.
\end{claim}
\begin{proof} Recall that the clique number of the graph,
namely the size of the largest clique in the graph, provides a lower
bound on the fractional chromatic number. Thus it suffices to show
that $\conf$ contains a clique of size $n+1$. Define the vertex set
$S = \{0\} \cup \{e_i\}_{i=1}^n$ of size $n+1$. For any $u,v \in S$,
$|u \oplus v|\leq 2$ and hence $S$ induces a clique in $\conf$
completing the proof of the inequality. The equality follows from the
previous claim. \end{proof}

\begin{corollary} \label{cor:ubound}
The parameters $\beta_1(H), \beta^*(H)$ satisfy
$\beta^*(H) = \beta_1(H) = \log_2 (n + 1)$.
\end{corollary}

\begin{proof}
Recall that $\beta_1(H)=\lceil\log_2\chi(\conf(H))\rceil$, and that in Theorem \ref{witsen-thm-1} we have actually shown that
$\beta^*(H) = \log_2 \chi_f(\conf(H))$. The proof therefore follows from the fact that $\chi(\conf) =\chif(\conf)=n+1$.
\end{proof}

Let $\conf_t = \conf_t(H)$ denote the confusion graph for
block-length $t$. Thus, $\conf_t$ is the Cayley graph of $\Z_2^{nt}$
whose generators are all vectors $\{(w_1,\ldots,w_n) ~|~ w_i \in
\Z_2^t ~,~ 1 \leq | \{i ~|~ w_i \neq 0\}| \leq 2\}$. In other words,
two vertices are connected in the confusion graph if they differ at
no more than 2 blocks.

\begin{claim} For $2^t \geq n$, $\chi(\conf_t) = \chi_f(\conf_t)=2^{2t}$.
\end{claim}
\begin{proof}
For a lower bound, it suffices to show a
set of size $2^{2t}$ which is a clique in $\conf_t$. Consider the
vertex set $\{(u_1,u_2,0,\ldots,0) ~|~ u_1,u_2 \in \Z_2^t\}$ in
$\conf_t$, which consists of $2^{2t}$ vertices. Every pair of
vertices in this set is connected in $\conf_t$ since they differ in
at most two blocks, and therefore this is a clique in $\conf_t$.
This shows that $\chi(\conf_t) \geq 2^{2t}$.

To complete the proof, we describe a proper coloring of $\conf$ which
uses $2^{2t}$ colors, using a simple Reed-Solomon code.
Let $\alpha_1,\ldots,\alpha_n$ be pairwise
distinct elements in the finite field $GF_{2^t}$, and define
the coloring $c: (GF_{2^t})^n \rightarrow GF_{2^t} \times GF_{2^t}$ as
follows. For a vector $u=(u_1,\ldots,u_n)$, let $c(u) :=
(\sum_{i=1}^{n} u_i~, \sum_{i=1}^{n} \alpha_i \cdot u_i)$. Clearly,
if $u,v \in (GF_{2^t})^n$ differ in exactly one block then the first
coordinate of $c(u)$ and $c(v)$ is different. Moreover, if $u$ and
$v$ differ in exactly two blocks $i,j$, then either $u_i + u_j \neq
v_i + v_j$ or $\alpha_i u_i + \alpha_j u_j \neq \alpha_i v_i +
\alpha_j v_j$ (or both inequalities hold), and again they will have
different colors as needed. This shows that the coloring $c$ is
indeed proper and completes the proof of the claim.
\end{proof}

Recalling that $\beta_t(H)=\lceil \log_2  \chi(\conf_t) \rceil$ we obtain the following
corollary, which together with Corollary \ref{cor:ubound} completes the proof of Theorem \ref{thm-dif-beta-beta-star}.

\begin{corollary} \label{cor:lbound} For the hypergraph $H$ defined above,
$\beta(H) = \lim_{t \rightarrow \infty}
\frac{1}{t} \log_2 \chi(\conf_t(H))
= 2 $.
\end{corollary}



\subsection[Proof of Theorem 1.3]{Proof of Theorem \ref{thm-dif-beta-star-beta-1}} \label{sec:gap_beta_star_beta_1}

The basic ingredient of the construction is a Cayley graph $G$ of an
Abelian group $K= \{k_1,\ldots,k_n\}$ of size $n$, for which there
is a large gap between the chromatic number and the fractional
chromatic number. In our context, we use $K=\Z_2^k$, though such
a Cayley graph of any Abelian group will do.
\begin{lemma}\label{lemma:kneser_ex}
For any $n=2^k$ there exists an explicit Cayley graph $G$ over
the Abelian group $\Z_2^k$
for which $\chi(G) > 0.01\sqrt{\log n}$ and yet $\chif(G) < 2.05$.
\end{lemma}
\begin{proof}

Let $n=2^k$, and consider the graph $G$ on the set of vertices $\Z_2^k$ as follows.
For any $i,j \in \Z_2^k$, the edge $(i,j) \in G$ if $|i \oplus j| \geq k -\frac{\sqrt k}{100}$.
We will now show that this graph has a large gap $\chi(G) / \chif(G) > \Omega(\sqrt{k}) = \Omega(\sqrt{\log n})$.
\begin{claim}
The chromatic number of $G$ satisfies $\chi(G) \geq
\frac{\sqrt{k}}{100}+2$.
\end{claim}
\begin{proof} The induced subgraph of $G$ on the vertices
$\{ i\in \Z_2^k ~|~ |i|=s \}$ where
$s=\frac{k}{2}-\frac{\sqrt{k}}{200}$, is the Kneser graph $K(k,s)$
whose chromatic number is precisely $k-2s+2$, as proved in
\cite{Lo}, using the Borsuk-Ulam Theorem. It is worth noting
that one can give a slightly simpler, self-contained
(topological) proof of this claim, based on the approach of \cite{Gre}.
\end{proof}

\begin{claim}
The fractional chromatic number of $G$ satisfies
$\chif(G) < 2.05$
\end{claim}

\begin{proof} Since $G$ is a Cayley graph, it is well
known that $\chif(G)=|V(G)|/\alpha(G)$ (c.f. e.g. \cite{SU}), and
therefore it suffices to show it contains an independent set of size
at least $\frac{2^k}{2.05}$. Let $I = \{ i \in \Z_2^k ~|~ |i| <
\frac{k}{2}-\frac{\sqrt k}{200} \}$. Obviously the set $I$ is an
independent set as the Hamming weight of $i \oplus j$ is below $k -
\sqrt{k}/100$ for any $i,j \in I$, and therefore $(i,j) \not \in
E(G)$. Hence,
\begin{equation*}
\chif(G) \leq \frac {2^k} {|I|}  =
\frac{2^k}{\sum_{i<\frac{k}{2}-\frac{\sqrt k}{200}} \binom{k}{i}} <
2.05 ~. \qedhere
\end{equation*}\end{proof}
This completes the proof of Lemma \ref{lemma:kneser_ex}.
\end{proof}

Let $H$ be the directed hypergraph on the
vertices $V=\{1,2,\ldots,n\}$ defined as follows. For each pair of
vertices $i,j$ such that $k_i,k_j$ are adjacent in $G$ (i.e.
$k_i-k_j$ is a generator in the defining set of $G$),
$H$ contains the directed edges
$(i, V \setminus \{i,j\})$ and $(j, V \setminus \{i,j\})$.
As before, every receiver misses precisely two blocks.

Let $\conf=\conf_1(H)$ be the confusion graph of $H$ (for block-length
$t=1$). Thus, $\conf$ is the Cayley graph of $\Z_2^n$ whose generators
are all vectors $e_i$, as well as all vectors $e_i \oplus e_j$ so
that $k_i,k_j$ are adjacent in $G$.

\begin{claim} \label{claim-integer-conf}
The chromatic number of $\conf$ satisfies $\chi(G) \leq \chi(\conf) \leq 3 \cdot
\chi(G)$.
\end{claim}

\begin{proof}
That fact that $\chi(G) \leq \chi(\conf)$ follows from the observation that the induced subgraph of $\conf$ on the
vertices $e_i$ is precisely $G$ (and hence, we similarly have $\chi_f(G) \leq \chi(\conf)$).

It remains to prove that $\chi(\conf) \leq 3 \chif(G)$. Let $c$ be some optimal coloring of $G$
with $d=\chi(G)$ colors. Define a coloring of $\conf$ with $3d$
colors as follows. For a vertex $x=x_1\ldots x_n$ assign the color
$c'(x)=\left(|x| \bmod 3,\sum_i x_i\cdot c(i)\right)$ where the sum is in $\Z_d$.
Clearly, $c'$ uses $3d$ colors. It remains to show that this is
indeed a legal coloring. Consider $x,y$ which are adjacent in
$\conf$, thus either $x\oplus y=e_i$ ($\Rightarrow |x|\not\equiv|y|
 \bmod 3$) or $x\oplus y=e_i \oplus e_j$. If $|x| \not\equiv
|y| \bmod 3$ they will have different colors, otherwise, $x \oplus y =
e_i \oplus e_j$, $|x|=|y|$ and there exists $z \in \Z_d$ so that
$c'(x)=(|x|,z+c(i))$ and $c'(y)=(|y|,z+c(j))$. Since $x,y$ are
adjacent, we know $i,j$ are adjacent in $G$, thus $c(i)\neq c(j)$
which implies that $c'(x) \neq c'(y)$ as required.

Note that in the special case where $G$ is a Cayley graph of the group $\Z_2^k$,
the above upper bound on $\chi(\conf)$ can be modified
 into the smallest power of $2$ that is strictly larger than $\chi(G)$.
\end{proof}

Notice that in the above claim we did use any of the properties of $G$,
hence they hold for any graph. This shows that regardless of
the choice of $G$, for the confusion graph defined above, the gap
between $\chi$ and $\chif$ is at most 3 times the corresponding gap in the original graph.

\begin{claim}\label{claim-fractional-conf}
The fractional chromatic number of $\conf$ satisfies
$\chif(G) \leq \chif(\conf) \leq 3 \cdot \chif(G)$.
\end{claim}

\begin{proof} As mentioned above, the lower bound on $\chif(\conf)$ follows from the
induced copy of $G$ in $\conf$, and it remains to show that $\chif(\conf) \leq 3 \chif(G)$.

Since $\conf$ is a Cayley graph, it
suffices to show it contains an independent set of size at least
$\frac{2^n}{3 \cdot \chif(G)}$. Let $I \subset K$ be a maximum
independent set in $G$. As $G$ is a Cayley graph, $\chif(G)=n/|I|$. For a vector $u=(u_1, \ldots ,u_n) \in \Z_2^n$, define $s(u)\in K$ by
$s(u)= \sum u_i \cdot k_i$. For any $j \in K$, put
$$
I_j=\{u \in \Z_2^n~|~ s(u) + j \in I \}.
$$
Let $u,v$ be a pair of vertices so that $u \oplus v = e_i \oplus e_j$
and $|u|=|v|$. Hence $s(u)=x+k_i$ and $s(v)=x+k_j$ (here we rely on
$K$ being Abelian). If $u,v$ both belong to $I_l$ for some $l \in
K$, it must be that $k_i$ and $k_j$ are not adjacent in $G$, and
thus $u$ and $v$ are not adjacent in $\conf$.

It now follows that if $u$ and $v$ are vectors in $I_j$ and $|u| \equiv |v| \pmod{3}$, then
$u$ and $v$ are not adjacent in $\conf$. Therefore, $I_j$ is a union
of three independent sets in $\conf$, and hence at least one of them
is of size at least $|I_j|/3$. This holds for every $j \in K$. When
$j \in K$ is chosen randomly and uniformly, then, by linearity of
expectation, the expected number of elements in $I_j$ is exactly
$$
2^n \cdot \frac{|I|}{n} = \frac{2^n}{\chif(G)}.
$$
Thus, there is some choice of $j$ for which $I_j$ is of size at
least $2^n/\chif(G)$, and $\conf$ contains an independent set of
size at least $|I_j|/3 = 2^n/3\chif(G)$, as needed.
\end{proof}

\begin{corollary}
For any Cayley graph $G$ of an Abelian group, there exists a
confusion graph $\conf$ and some $c \in [\frac{1}{3},3]$ such that $$\frac {\chi(\conf)} {\chif(\conf)} = c
\cdot \frac {\chi(G)} {\chif(G)}~.$$
\end{corollary}

Plugging the graph which is guaranteed by Lemma
\ref{lemma:kneser_ex} into this construction completes the proof of
Theorem \ref{thm-dif-beta-star-beta-1}.


\begin{remark}
If we set $G = K_n$ which is indeed a Cayley graph over an Abelian
group, we get the example from Section \ref{sec:gap_beta_beta_star}.
Some of the claims in this section generalize claims from Section
\ref{sec:gap_beta_beta_star}.
\end{remark}

\subsection{Applications to Index and Network Coding} \label{sec:disjoint_union}

We are now considering the more restricted model where there is a
single receiver which is interested in each block. In the directed
hypergraph notation, this is equivalent to having precisely $m=n$
directed edges where each directed edge has a different origin
vertex. For easier notations, we can describe such a scenario (as
done in \cite{BBJK}, \cite{LS}) by a directed graph. Each directed
edge $(i,J)$ will be translated into $|J|$ directed edges $(i,j)$
for all $j\in J$. We can also consider an undirected graph for the
case where the receiver who is interested in $x_i$ knows $x_j$ iff
the receiver who is interested in $x_j$ knows $x_i$. We use similar
notations to the directed hypergraphs.

Clearly, $\beta_1(k \cdot G) \leq k \cdot \beta_1(G)$, as one can
always obtain an index code for $k \cdot G$ by taking the $k$-fold
concatenation of an optimal index code for $G$. Furthermore, it is
not difficult to see that this
bound is tight for all perfect graphs. Hence, the smallest graph
where $\beta_1(k \cdot G)$ may possibly be smaller than $k \cdot
\beta_1(G)$ is $C_5$, the cycle on $5$ vertices - the smallest
non-perfect graph. Indeed, in this case index codes for $k \cdot
C_5$ can be significantly better than those obtained by treating
each copy of $C_5$ separately. This is stated in Theorem \ref{witsen-thm-k-C5}
which we now prove.

%
\begin{proof}[\textbf{\emph{Proof of Theorem \ref{witsen-thm-k-C5}}}]
One can verify that the following is a maximum independent set of
size $5$ in the confusion graph $\conf(C_5)$: $$\left\{ 00000, 01100, 00011, 11011,
11101 \right\}~.$$ In the formulation of Theorem
\ref{witsen-thm-1}, $\gamma=5$, and this theorem now implies that
$\beta_1(k \cdot C_5) / k$ tends to $5-\log_2 5$ as $k\to\infty$.
On the other hand, one can verify\footnote{This fact can be
verified by a computer assisted proof, as stated in \cite{BBJK}.}
that $\chi(\conf(C_5)) = 8$, hence $\beta_1(C_5)=3$.
\end{proof}

This shows that there is a graph $G$ with an optimal index code $\cal
C$, so that much less than $|{\cal C}|^k$ words suffice to
establish an index code for $k\cdot G$, although each of the $k$
copies of $G$ has no side information on any of the bits
corresponding to the remaining copies.


\begin{remark}
  Using the upper bound of \eqref{witsen-eqcode-bound}
  in its alternate form, as stated in Remark
  \ref{witsen-rem-lov}, we obtain that $\beta_1(k \cdot C_5) < k \cdot \beta_1(C_5)$
  already for $k=15$.
\end{remark}

  The example of $C_5$ can be extended to other examples by looking at all the complement of odd cycles, i.e. $\overline{C}_k$ for any odd $k \geq 5$. All graphs in this family have a gap between the optimal code for disjoint union in comparison to the concatenation of the optimal code for a single copy. In Appendix \ref{sec:odd-antiholes} we prove the following properties of the complements of odd cycles:

    \begin{claim}\label{constant_c_odd_antiholes}
There exists a constant $c > 1$ so that for any $n \geq 2$,
$\chi(\conf(\overline{C}_{2n+1})) > c \cdot \chif(\conf(\overline{C}_{2n+1}))$.
\end{claim}
\begin{theorem}\label{thm-odd-cycle-complements}
Let $H_{2n+1} = ([2n+1],E)$, where for each $i \in [2n+1]$
there is a directed edge $(i, N_{\overline{C}_{2n+1}}(i))$ in $E$,
and $N_{\overline{C}_{2n+1}(i)}$ are the neighbors of $i$ in $\overline{C}_{2n+1}$. Then any linear
code for $H_{2n+1}$ requires 3 letters. \end{theorem}
Theorem \ref{thm-odd-cycle-complements} implies that a
broadcast network based on the complement
of any odd cycle has linear code of minimal length 3, regardless of the
block length. Specifically for $\overline{C}_{23}$, we know that
$\chif(\conf(\overline{C}_{23})) \leq 4.809$ (as can be seen in
Appendix \ref{sec:odd_holes_numbers}). Therefore, the above mentioned reduction to Network Coding provides
us with an explicit network (of size 48) where the linear code must be of
length at least 3 whereas the optimal code can be of length $\beta
\leq \beta^* \leq \log_2 4.809 \approx 2.265$, yielding a ratio of
$1.324$. This proves Corollary \ref{network-code-ratio}. \qed


\section{Conclusions and open problems} \label{sec:conclusion}
\begin{itemize}
  \item In this work, we have shown that for every broadcast network $H$ with $n$ blocks
  and $m$ receivers, and for large values of $k$,
  $\beta^*_k(H) = \beta_1(k \cdot H) = \left( n - \log_2
      \alpha(\conf_1(H)) +o(1)\right)k$, where the
      $o(1)$-term tends to $0$ as $k\to\infty$.
For every large constant $C$ there are examples $H$ such that for
large $k$, $\beta^*_k(H)/k<3$ and yet $\beta_1(H)>C$.


  \item Our results also imply that encoding the entire block at once can be strictly
  better than concatenating the optimal code for H with a single bit block.
   This justifies the definition of the broadcast rate of $H$, $\beta(H)$, as the optimal
   asymptotic average number of bits required per a single bit of
coding in each block for $H$.

  \item We have shown an infinite family of graphs (including the
  smallest possible non-perfect graph $C_5$) for which there exists
  a constant $c>1$ so that for each of these graphs there is a multiplicative
  gap of at least $c$ between the chromatic number and
  the fractional chromatic number of their confusion graphs. However,
  the gap in all these graphs is below 2,
and it is  not known  if for graphs this gap can be arbitrarily large.

  \item Generalizing the above setting, allowing multiple users to request
  the same block, allows us to construct hypergraphs whose confusion
  graphs exhibit bigger gaps.

  \begin{itemize}
    \item We have shown a specific family of confusion graphs where
  the fractional chromatic number is bounded ($<7$) while the
  chromatic number is unbounded ($\Omega(\sqrt {\log n})$). In these
  settings, a 1-bit block-length will require us to transmit $\Theta(\log\log n)$
  bits while for large $t$-bit block-length, the required number of bits
   is linear in $t$. For other families, this ratio can even reach $\Theta(\log n)$. More surprisingly, for the first family, a network
consisting of $t$ independent copies of the original one will only
require a number of bits that is linear in $t$.
    \item With the generalized construction, one can build a hypergraph for any
    Cayley graph of an Abelian group for which the confusion graph maintains the
    same gap as the original graph. The maximum gap that can be obtained in
    this way is $O(\log n)$, since this is the maximum possible gap
   between the fractional and integer chromatic numbers of any $n$-vertex graph
    (c.f., e.g.,\cite{SU}).
  \end{itemize}

  \item Currently, there is no known better upper bound for this gap which
  is specific for confusion graphs. The examples above are all
  exponentially far from the general upper bound $\Theta(\log V)$,
  which in our case equals to $\Theta(\log 2^n) = \Theta(n)$.

  \item An interesting  problem in Network Coding is that of
  deciding whether or not there are
  networks with an arbitrarily large gap between
the optimal linear and
  non-linear flows. Note that the network is
  not allowed to depend on the size of the underlying field.
  Generalizing our constructions to create such examples could be
  interesting.

\end{itemize}


\newpage
\appendix
\section{Appendix}

\newcommand{\GG}{\conf_{2n+1}}
\newcommand{\GGG}{\conf_{2n+3}}

\subsection{Proof of Theorem \ref{thm-odd-cycle-complements}}
We first need to present
some definitions and theorems from \cite{BBJK}. We say that a matrix $A$ {\em
fits} a graph $G = (V,E)$ if $A[i,i]\neq 0$ for all $i$ and $A[i,j]=0$ for
$i\neq j ~,~ (i,j) \not\in E$ (for $(i,j) \in E$, $A[i,j]$  is not limited).
A generalization of a result in \cite{BBJK} (as noted in \cite{LS}) states
that the length of the minimal linear encoding of $G$ over a field $\mathbb{F}$
is always at least the minimal rank over $\mathbb{F}$ of a matrix $A$ which fits $G$.
It therefore suffices to show the following:

\begin{claim} \label{linear_lower_bound} Let $A$ be a matrix that fits
the graph $\overline{C}_{2n+1}$ over some field $\mathbb{F}$. Then $\rank(A) \geq 3$.
\end{claim}

\begin{proof}
Let $A$ be a matrix that fits $\overline{C}_{2n+1}$.
This means that
\begin{align*}
& \forall i ~.~ A[i, i] \neq 0 ~,\\
& \forall i \in [2n] ~.~ A[i,i+1] = 0 ~,\\
& \forall i \in [2n] ~.~  A[i+1,i] = 0 ~,\\
& A[1,2n+1] = A[2n+1,1] = 0~.
\end{align*}
Let $A(t)$ denote the $t$'th row of $A$, when we look at it as a
vector. Note that $A(1), A(2)$ are linearly independent, as $A[1,1]
\neq 0$ but $A[2,1] = 0$ and $A[2,2] \neq 0$. Assume, towards a
contradiction, that $\rank(A) = 2$ and hence for every $t$:  $A(t) =
a_t A(1) + b_t A(2)$. We prove by induction on $t$ that if $t$ is
odd then $b_t = 0$, and if $t$ is even then $a_t = 0$. Note that for
any  $t$ it is impossible that $a_t = b_t = 0$, as each line has a
nonzero element. For $t = 1,2$ this is trivial. For some odd $t = 2k
+1$, by assumption $A(2k) = b_{2k} A(2)$ and $A(2k-1) = a_{2k-1}
A(1)$. This means that
\[A(2k+1) =
a_{2k+1}/a_{2k-1} A(2k-1) + b_{2k+1}/b_{2k}A(2k)\] As $A[2k-1,2k] =
A[2k+1,2k] = 0$ but $A[2k,2k] \neq 0$, this means that $b_{2t+1} =
0$, as required. A similar argument works for even $t$, which
completes the proof of the induction. However, this leads to a
contradiction, as $A(2n+1) = a_{2n+1} A(1)$, and this is impossible
as $A[2n+1,1]=0$ but $A[1,1] \neq 0$. Altogether, the assumption
that $rank(A)=2$ leads to a contradiction, hence $\rank(A) \ge 3$.
\end{proof}

Claim \ref{linear_lower_bound} completes the proof of Theorem \ref{thm-odd-cycle-complements}.

\subsection{Complements of odd cycles} \label{sec:odd-antiholes}

We showed that the cycle of length 5 is the
smallest graph where there exists a gap between the fractional and
integer chromatic numbers of its confusion graph. The cycle and its
complement on 5 vertices are isomorphic,
however this is not the case for larger odd cycles and their
complements. We now
show that there is a gap between those numbers for any complement of an odd cycle of 5
or more vertices.

Throughout this section, let $\GG = \conf(\overline{C}_{2n+1})$.

\begin{claim}
Any independent set $A$ of $\GG$ can be extended to an
independent set $A'$ in $\GGG$ where $|A'| = 4|A|$.
\end{claim}

\begin{proof} We first define a function $f$ from the
vertices of $\GG$ into sets of size 4 from the vertices of
$\GGG$ which satisfies the following:
\begin{itemize}
  \item For every vertex $v$ of $\GG$, $f(v)$ is an independent set in $\GGG$.
  \item If $u$ and $v$ are not adjacent in $\GG$, then $f(v) \cup f(u)$ is an independent set in $\GGG$.
  \item $f(v) \cap f(u) = \emptyset$ for any $u \neq v$.
\end{itemize}

Given such $f$ and an independent set $A$ of $\GG$,
define $A' = \bigcup_{v \in A} f(v)$ which will be an independent
set of size $4|A|$ in $\GGG$ as needed. We now describe this $f$
explicitly and prove its properties:

$$f(v=(x_1,x_2,\ldots,x_{2n},x_{2n+1})) =  \left\{\begin{array}{c}
  (x_1,x_2,\ldots,x_{2n},x_{2n+1},0,0)                       = v'\oplus m_0 \\
  (x_1,x_2,\ldots,x_{2n},\overline{x_{2n+1}},0,1)            = v'\oplus m_1 \\
  (\overline{x_1},x_2,\ldots,x_{2n},x_{2n+1},1,0)            = v'\oplus m_2 \\
  (\overline{x_1},x_2,\ldots,x_{2n},\overline{x_{2n+1}},1,1) = v'\oplus
  m_3
\end{array}\right\}$$
where $v'$ is $v$ extended to length $2n+3$ with two additional
zeros at the right end and $m_0,m_1,m_2,m_3$ are $4$ appropriate constant binary
vectors of length $2n+3$.

\begin{itemize}
  \item $f(v)$ is an independent set: Since $\GGG$ is a Cayley graph
  over $\Z_2^{2n+3}$,
  we only need to show that the sums of all pairs of the 4 vectors
  in $f(v)$ are not generators in our graph. All these sums are in
  $\{m_1,m_2,m_3\}$ since $m_0=0$ and $m_1\oplus m_2 \oplus m_3=0$
  (notice that these sums are independent of the choice of
  $v$).
  Since we are looking at the confusion graph of the complement of an odd cycle, the
  generators of this graph are vectors of hamming weight 1,2 and 3 of
  consecutive ones (i.e. $e_i$, $e_i\oplus e_{i+1}$ and
  $e_i \oplus e_{i+1} \oplus e_{i+2}$ for any
  $i$ where the indices are reduced modulo $2n+3$). Indeed
  $\{m_1,m_2,m_3\}$ are not generators, therefore $f(v)$ is independent.

  \item $f(v) \cup f(u)$ is an independent set when $u$ and $v$ are not adjacent:
  Consider $x = v' \oplus m_i$ and $y = u' \oplus
  m_j$ for some $i,j$ (where $u'$ and $v'$ are as before). We want to show that $x \oplus y$ is not a
  generator. If $i=j$ then $x \oplus y = u' \oplus v'$ and therefore
  it is not a generator (two additional zero bits at the right will
  not turn a non-generator into a generator). Assume now $i \neq j$, then we know $x\oplus
  y = (u' \oplus v') \oplus m_k = (u \oplus v)' \oplus m_k$ for some $k \in \{1,2,3\}$. Since
  $u \oplus v$ is not a generator, with claim \ref{not_gen_xor_mask}
  we get that $x \oplus y$ is not a generator as needed.

  \item For $v \neq u$,  $f(v) \cap f(u) = \emptyset$: Let us assume
  there exists some $x \in f(v) \cap f(u)$. Since all vectors in both
  $f(v)$ and $f(u)$ differ at their two right most bits $\{x_{2n+2}, x_{2n+3}\}$,
  there exists $i \in \{0,1,2,3\}$ so that $x=v' \oplus m_i = u' \oplus m_i$
  in contradiction to $v \neq u$. \qedhere
  \end{itemize}
\end{proof}

\begin{claim}\label{not_gen_xor_mask}
If $x$ of length $2n+1$ is not a generator, then $x' \oplus m_k$ for
$k \in \{1,2,3\}$ is not a generator as well (where $x'$ is $x$
extended with two zero bits on the right as before).
\end{claim}

\begin{proof} The cases of $k=1$ and $k=2$ are symmetric
so we will consider only $k=1$ and $k=3$. We show explicitly what
$x$ can be in order for the result to be a generator, and as all such
$x$ vectors turn out to be generators the desired result follows. Since $x'$ is
the extension of $x$ with two zero bits on the right, it cannot
affect the two rightmost bits of $x' \oplus m_k$.
\begin{itemize}
  \item $k=1$: $m_1 = 0 \overbrace{0\ldots 0}^{2n-1} 101$ so in
  order to make it a generator, we must flip the $2n+1$ bit and then
  we can at most flip 0,1 or 2 consecutive ones at the left most
  side. Hence,
  $x \in \{
   00 \underbrace{0\ldots 0}_{2n-2} 1,
   10 \underbrace{0\ldots 0}_{2n-2} 1,
   11 \underbrace{0\ldots 0}_{2n-2} 1
\}$ which are all generators of length $2n+1$.

  \item $k=3$: $m_3 = 1 \overbrace{0\ldots 0}^{2n-1} 111$ so in
  order to make it a generator, we must flip at least one of the bits
  at locations $\{1,2n+1\}$ and no other bit. Thus, \\
  $x \in \{
   0 \underbrace{0\ldots 0}_{2n-1} 1,
   1 \underbrace{0\ldots 0}_{2n-1} 0,
   1 \underbrace{0\ldots 0}_{2n-1} 1
\}$ which are all generators of length $2n+1$.\qedhere
\end{itemize}
\end{proof}

\begin{claim}\label{claim-frac-decreasing}
The fractional chromatic number of $\GG$ is monotone decreasing
with $n$, that is, $\chif(\GGG) \leq \chif(\GG)$.
\end{claim}

\begin{proof} Since $G$ is a Cayley graph, we know
$\chif(\GG) = 2^{2n+1} / \alpha(\GG)$. Using the previous
claim we know $\alpha(\GGG) \geq 4\alpha(\GG)$ and
therefore \begin{equation*}
\chif(\GGG) = 2^{2n+3} / \alpha(\GGG) \leq
2^{2n+1} / \alpha(\GG)~.\qedhere
\end{equation*} \end{proof}

\begin{corollary}
For any $n\geq 8$, $\chif(\GG) < 4.99~(~< 5)$.
\end{corollary}

\begin{proof} By a computer search (see Appendix \ref{sec:odd_holes_numbers}) one can
see that the fractional chromatic number of the confusion
graph of the complement of an odd cycle on 17 vertices is below 5. Since this property
is monotone, this holds for any $n \geq 8$. \end{proof}


\begin{proof}[\textbf{\emph{Proof of Claim \ref{constant_c_odd_antiholes}}}] The authors of \cite{BBJK} showed that the
chromatic number of the confusion graph of any complement of an odd cycle
is strictly bigger than 4 and is
at most 8. We have shown that the fractional
chromatic number of the confusion graph of a cycle on 5 vertices is
32/5=6.4
and we just proved it is monotone decreasing. Since the number of
vertices in these confusion graphs is a power of 2, the fractional
chromatic number cannot be an integer between 4 and 8 so it will
always be smaller than the integer chromatic number.

We know that for any $n\geq 8$ there is a gap of at least
$5/\chif(\conf_{17})$. For smaller values of $n$ there is some fixed gap
exceeding $1$, so
taking the minimum between these gaps will give a single $c$ for
which the claim holds. \end{proof}

The limit $\lim_{n\rightarrow\infty} \chif(\GG)$, which
exists by monotonicity, remains unknown at this time.

\begin{claim}
The chromatic number of $\GG$ is monotone decreasing with $n$:
 $\chi(\GGG) \leq \chi(\GG)$.
\end{claim}

\begin{proof} A coloring of $\GG$ with $k$ colors is
a partition of the graph into $k$ independent sets. We can obtain a
coloring of $\GGG$ with the same number of colors by applying
the extension described before on each of the independent sets. We
already proved the new sets would be independent sets in the new
graph. We need to prove that all the vertices of the graph belong to
one of the independent sets. This can be shown by noticing the size
of the new sets is 4 times their previous size and since they have
empty intersection (as we have seen before) they must cover the
entire graph (as the size of the graph is precisely 4 times
that of the previous one).  \end{proof}

\begin{corollary}
For any $n\geq 3$, $5 \leq \chi(\GG) \leq 7$.
\end{corollary}

\begin{proof} By a computer search (see Appendix \ref{sec:seven_coloring})
one can
see that for 7 vertices, the integer chromatic number is at
most 7. By monotonicity and the fact that is must be at
least 5 (as shown in \cite{BBJK})  the desired result follows.
\end{proof}

As in the fractional case, the limit $\lim_{n\rightarrow\infty}
\chi(\GG)$ exists and can only be 5,6 or 7, however it remains
unknown at this time.

%
%

\subsection{Fractional chromatic number upper bounds for
$\conf\left(\overline{C}_n\right)$} \label{sec:odd_holes_numbers}

Here is a table of upper bounds for the fractional chromatic number
of the confusion graphs of the complements of odd cycles. These upper
bounds were found by searching for a large independent set in each
of these graphs as they are Cayley graphs. This search was done by a
computer program which does not assure us for the optimal result,
hence it only provides the bounds stated in the table, which are
not necessarily tight.

$$\begin{array}{c|c|c}
    \ \ \ \ n \ \ \ \ & \ \ \ \ \alpha(\conf\left(\overline{C}_n\right)) \ \ \ \ & \chif(\conf\left(\overline{C}_n\right)) \\
  \hline
  5 &           5 &  2^5/5                = 6.4 \\
  7 & \geq         22 & \leq 2^7/22               \approx 5.818 \\
  9 & \geq         93 & \leq 2^9/93               \approx 5.505 \\
 11 & \geq        386 & \leq 2^{11}/386           \approx 5.306 \\
 13 & \geq       1586 & \leq 2^{13}/1586          \approx 5.165 \\
 15 & \geq       6476 & \leq 2^{15}/6476          \approx 5.060 \\
 17 & \geq      26317 & \leq 2^{17}/26317         \approx 4.981 \\
 19 & \geq     106744 & \leq 2^{19}/106744        \approx 4.912 \\
 21 & \geq     430592 & \leq 2^{21}/430592        \approx 4.870 \\
 23 & \geq    1744414 & \leq 2^{23}/1744414       \approx 4.809
\end{array}$$

Although the bounds are not necessarily tight, they clearly
suggest a monotone behavior of the fractional
chromatic numbers of these graphs.

The computer program which found most of these sets and a program
that verifies an independent set (of a specific form) can be found
at \protect\url{www.math.tau.ac.il/~amitw/broadcasting}.

\subsection{Coloring the confusion graph of $\overline{C}_7$
with 7 colors}\label{sec:seven_coloring}

It was proved in \cite{BBJK} that the index coding for any complement of an odd cycle
is precisely 3, however, the minimum number of codewords can vary
between 5 and 8. Here we show a legal coloring using 7 colors for
$n=7$, which was found using a computer program. Each cell in the
following table represent a vertex out of the 128 vertices in the
graph which are $\{0,1,\ldots 127\}$. The vertex is the sum of its
two indices in the table, e.g. the bolded vertex in the table is
$16+4$ which is colored with the seventh color.

$$\begin{array}{c|cccccccccccccccc}
 & \ 0 \ & \ 1 \ & \ 2 \ & \ 3 \ & \ 4 \ & \ 5 \ & \ 6 \ & \ 7 \ & \ 8 \ & \ 9 \ & 10 & 11 & 12 & 13 & 14 & 15  \\
 \hline
 0  & 1 & 2 & 3 & 4 & 2 & 7 & 6 & 5 & 3 & 4 & 7 & 2 & 4 & 1 & 5 & 7  \\
 16 & 2 & 7 & 4 & 3 & \mathbf{7} & 5 & 1 & 4 & 4 & 3 & 2 & 1 & 5 & 6 & 7 & 2  \\
 32 & 3 & 4 & 7 & 2 & 4 & 1 & 5 & 7 & 1 & 2 & 3 & 4 & 2 & 7 & 6 & 5  \\
 48 & 4 & 3 & 2 & 1 & 5 & 6 & 7 & 2 & 2 & 7 & 4 & 3 & 7 & 5 & 1 & 4  \\
 64 & 5 & 7 & 1 & 6 & 3 & 5 & 2 & 4 & 7 & 6 & 5 & 1 & 6 & 3 & 4 & 2  \\
 80 & 1 & 5 & 6 & 7 & 2 & 4 & 3 & 6 & 6 & 1 & 7 & 5 & 4 & 2 & 5 & 3  \\
 96 & 7 & 6 & 5 & 1 & 6 & 3 & 4 & 2 & 5 & 7 & 1 & 6 & 3 & 5 & 2 & 4  \\
 112& 6 & 1 & 7 & 5 & 4 & 2 & 5 & 3 & 1 & 5 & 6 & 7 & 2 & 4 & 3 & 6
\end{array}$$

A computer program that verifies this is indeed a legal coloring can
also be found at \\
\protect\url{www.math.tau.ac.il/~amitw/broadcasting}.

\end{document}